# Optical Shaping of Plasma Cavity for Controlled Laser Wakefield Acceleration

Bobbili Sanyasi Rao[1,2†], Myung Hoon Cho[1,3], Hyung Taek Kim[1,4#], Jung Hun Shin[1], Kyung Hwan Oh[1], Jong Ho Jeon[1], Byung Ju Yoo[1], Seong Ha Cho[1], Seong Ku Lee[1,4], Chang Hee Nam[1,5]

[1]*Center for Relativistic Laser Science (CoReLS), IBS, Gwangju 61005, Korea*
[2]*Laser Plasma Division, Raja Ramanna Centre for Advanced Technology, HBNI, Indore 452013, India*
[3]*Pohang Accelerator Laboratory, Pohang, Gyeongbuk 37673, Korea*
[4]*Advanced Photonics Research Institute, GIST, Gwangju 61005, Korea*
[5]*Department of Physics and Photon Science, GIST, Gwangju 61005, Korea*
[†]*sunnyb@rrcat.gov.in / [#]htkim@gist.ac.kr*

Laser wakefield accelerators rely on relativistically moving micron-sized plasma cavities that provide extremely high electric field >100GV/m. Here, we demonstrate transverse shaping of the plasma cavity to produce controlled sub-GeV electron beams, adopting laser pulses with an axially rotatable ellipse-shaped focal spot. We showed the control capability on electron self-injection, charge, and transverse profile of the electron beam by rotating the focal spot. We observed that the effect of the elliptical focal spot was imprinted in the profiles of the electron beams and the electron energy increased, as compared to the case of a circular focal spot. We performed 3D particle-in-cell (PIC) simulations which reproduced the experimental results and revealed dynamics of a new asymmetric self-injection process. This simple scheme offers a novel control method on laser wakefield acceleration to produce tailored electron beams and x-rays for various applications.

The laser wakefield acceleration (LWFA) has attracted attention intensively since its conception [1-3] as a potential alternative to the conventional technology for developing a compact TeV electron-positron collider [4] and ultrashort bright x-ray sources [5]. Recently, nearly 10 GeV electron beams has been demonstrated through LWFA from 10 cm-scale plasma with a petawatt laser pulse [6, 7]. Further, significant efforts are devoted to the efficient control of the injection and acceleration processes to produce controllable high-quality electron beams [8-16] and bright radiation sources [17-21] for a variety of applications.

LWFA exploits strong wakefields (≥100 GV/m) in plasma excited by the ponderomotive force ($F_p$) of an intense laser pulse. The ponderomotive force ($F_p \propto -\nabla I_L$, where $I_L$ is laser intensity) of a focused short-pulse laser displaces plasma electrons in all directions, while heavier ions stay behind. The charge separation forms a spherical plasma wave, so-called plasma cavity or bubble, when the pulse-length and focal spot radius of the laser match half the plasma wavelength [22,23]. At sufficiently high intensity the laser pulse induces nonlinear plasma wave breaking which causes self-injection of electron bunch into the plasma cavity from its back-end [24]. The injected electrons experience accelerating and focusing forces due to axial and transverse components of the wakefield in the plasma cavity, respectively. The radial force acts to transversely confine the accelerating electrons and causes their transverse oscillations, known as "betatron oscillations" around the laser axis, which produces highly directional ultrashort (~fs) synchrotron radiation [21].

When a laser pulse is focused to a perfect circular spot, the associated ponderomotive force produces a transversely symmetric plasma cavity and wakefield. Studies on electron injection in LWFA [25,26] and emission of betatron radiation [19] mostly consider spherical plasma cavities. In principle, shaped laser focal spot can tailor the transverse shape of the plasma cavity for delicate control of electrons' self-injection and their trajectories which can then facilitate generation of controlled electron beams and radiation. Earlier, the effects of off-centered intensity profile or an astigmatic aberration of drive laser pulse have been studied to manipulate betatron oscillations and radiation [20, 27-29]. However, direct control of the LWFA and transverse properties of the electron beam by shaping the laser focus have not been shown so far, even though the tuning of LWFA by manipulating the plasma cavity can play crucial role in better understanding and future development of LWFA.

We report here demonstration of stable sub-GeV quasi-monoenergetic electron beams with enhanced energy and tailored transverse profile from LWFA driven by a laser pulse with an ellipse-shaped focal spot. We show that the orientation, ellipticity, and charge of the transversely ellipse shaped electron beam could be tuned by simply rotating the focal spot. We proved in experiments and PIC simulations that the shaped laser focus could be effectively used for shaping the plasma cavity and fine control of self-injection in LWFA, which can be exploited for producing tunable electron and x-ray beams in future.

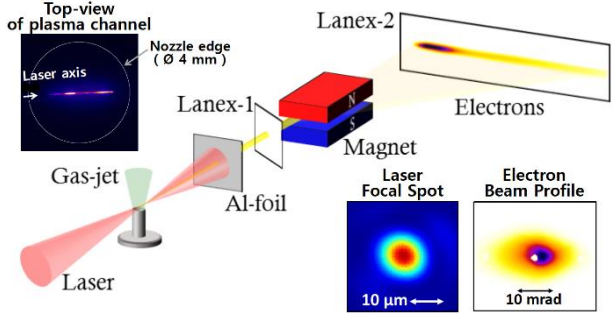

Figure 1. Schematic of the experimental set-up used for laser wakefield acceleration. The images of the laser focal spot and the corresponding electron beam profile without shaping of the laser beam are shown in the inset at bottom-right. The top-view image of the plasma channel is shown in the inset at top-left. The focal spot was nearly circular after corrected with an adaptive optics.

The experiment was performed using a femtosecond Ti:sapphire laser with a peak power of 150 TW. The laser delivered horizontally polarized ultrashort laser pulses of full-width-at-half maximum (FWHM) duration, $\tau_L$ = 27±2fs. The transverse profile of the laser beam before focusing was circular and had a flat-top intensity distribution with measured FWHM diameter 70 mm. A schematic of the experimental setup is shown in Fig. 1. The laser beam was focused, using a spherical focussing mirror with f-number 17, into a helium gas jet produced from a cylindrical nozzle with an orifice diameter of 4 mm. The density profile of the gas jet was measured using interferometry. The laser focus was kept at a height of 1 mm from the nozzle where the gas jet has 2-mm wide flat-top density profile with 1mm up- and down-ramps. The laser propagation and its interaction length in the helium gas jet plasma were monitored through top-view imaging of the plasma radiation.

The electron beams produced by the LWFA was characterized using the diagnostics set up in the laser downstream direction. A thin aluminium foil (Al-foil), kept after the interaction, blocks the residual drive laser beam. The scintillating screens (Lanex™) Lanex-1 (kept at a distance of 35cm from the gas jet) and Lanex-2 (kept at a distance of 52cm from the exit of the magnet) were covered on front with 0.1mm thick Al-foils to cut off low energy background electrons and rear sensitive surfaces are imaged with CCD cameras for monitoring electron beam before and after the magnet. The images obtained from the Lanex-1 were used to measure the pointing, transverse shape, divergence, and total charge of the electron beam. A permanent dipole magnet, having 0.96T over 20.5cm length within a pole gap of 8 mm, was used to measure the electron energy spectrum. The Lanex-2 screen was set to detect electrons in the energy range 0.1 - 0.5 GeV. The resolution of electron energy measurement was 2 % at 0.1 GeV and 16% at 0.5 GeV by considering the beam divergence of 10 mrad.

We have employed an adaptive optics system to correct the wavefront distortion of the laser beam to avoid the effects of wavefront aberration and obtain a high-quality focal spot. A typical image of measured focal spot is shown in an inset of the Fig. 1. The focal spot was nearly circular with average $w_0$ of 17 µm. For the focal-spot shaping, we introduced an ellipse-shaped hard aperture in the laser beam path centred on the beam axis, prior to sending the laser beam onto the focusing mirror. The elliptic aperture has a diameter of 65 mm × 43.3 mm and ellipticity of 1.5. The aperture could be translated in and out of the beam path and could also be rotated about the axis of the laser beam. The introduction of the aperture caused the laser beam to acquire ellipse shape and transmitted nearly 60% of the laser energy.

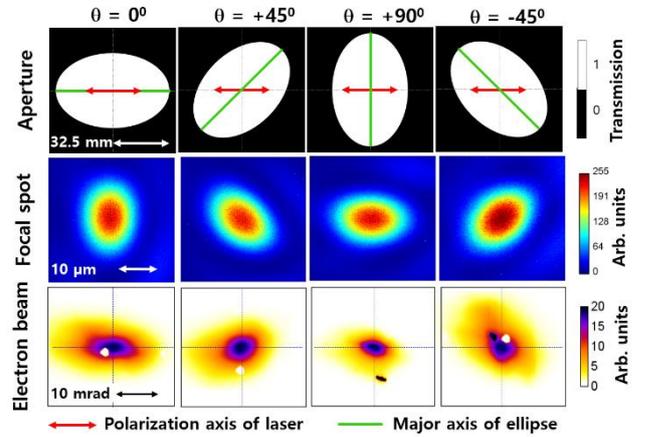

Figure 2. Top row - representative images of the ellipse-shaped hard aperture oriented in four different directions w.r.t. the laser polarization direction, and the corresponding images of the laser focal spot and the typical electron beam profile recorded when the aperture was introduced in the drive laser beam. The white dots in the electron beam images are reference markers on Lanex screen used for the calibration of imaging.

The focal spot images of the ellipse-shaped laser beam were measured for different orientations of the ellipse w.r.t. the laser polarization axis, as shown in Fig. 2. The focal spot was clearly elliptical with its major axis aligned perpendicular to that of the elliptical aperture. The rotation of the focal spot with the rotation of aperture could also be noticed in the figure. The $1/e^2$ radius of the shaped focal spot was measured to be 25 µm × 17 µm which gives rise to the ellipticity, $e \approx 1.5$, where we define ellipticity $e = a/b$; $a$ and $b$ are respectively the major and the minor radius of the ellipse. The peak laser intensity, $I_L$ in the focus is $8.4 \times 10^{18}$ W/cm$^2$ and the corresponding normalized vector potential, $a_0 = 0.85[I_L(\times 10^{18}$ W/cm$^2).\lambda^2(\mu m)]^{1/2} \approx 2$, considering the transmission of the aperture and the energy concentrated in the focal spot. Corresponding to the 60% transmission of the aperture, the laser beam without aperture provided higher peak intensity of $1.7\times10^{19}$ W/cm$^2$ and $a_0 \approx 2.5$.

The LWFA experiment was initially carried out without the elliptic aperture and reproducible electron beams were obtained from self-injected LWFA at plasma density, $n_e \approx 1\times10^{19}$cm$^{-3}$ ($c\tau = 0.8\lambda_p > \lambda_p/2$) from 2.3±0.1mm long plasma channel. The laser power was reduced to about 60 % of maximum power in order to match the laser energy on target with and without aperture. The electron beam measurements were performed in 20-30 laser shots for each set of experimental conditions and obtained the mean and the standard error [30] of the mean for the relevant beam parameters. Images of the typical electron beam profile and the laser-plasma channel are shown in the insets of Figure 1. In spite of the circular focal spot, the electron beam profile elongated along the laser polarization axis, indicating the overlap of accelerating electrons with the drive laser field inside the back-half of the bubble [29]. The direct influence of the laser field was further evident during the experiment from the observation of larger ellipticity at higher plasma density and/or laser power. The FWHM contour of the electron beam profile was fitted to a closest ellipse to quantify the ellipticity of the electron beam. The electron beam has divergence angles 10.2±0.7mrad and 6.0±0.4 mrad along the horizontal and the vertical planes, respectively. The ellipticity of the electron beam was 1.74±0.10 and the orientation of its major axis was -5.2$^0$±3.8$^0$. The root-mean-square (rms) pointing fluctuations of the electron beams was 1.9 mard and 1.7 mard in both the planes.

To demonstrate the effect of ellipse-shaped focal spot on LWFA, the elliptical aperture was inserted in the laser beam. The images in the third row of Figure 2 show the typical profiles of electron beams recorded for different orientations of the focal spot w.r.t. the laser polarization axis. Figure 2 clearly indicates that the major axis of the electron beam follows the rotation angle of the aperture except for the case of θ=90$^0$. In the case of θ=90$^0$, the major axis of the elliptic electron beam possibly could not be aligned with that of the aperture due to the strong effect of the laser polarization along the horizontal axis.

The energy spectra of the electron beams were compared for the two cases obtained with and without the aperture in the laser beam path for the same laser energy on target. We found that the shaped laser focus can stabilize the electron injection resulting in electron beams with better reproducibility. The averaged energy spectra with rms errors measured with and without the aperture are shown in Fig. 3. The electron beams with the elliptic focal spot had enhanced stability and higher peak energy compared to the circular focus result. We could not observe considerable change in the peak energy of the electron beam with the rotation of the focal spot.

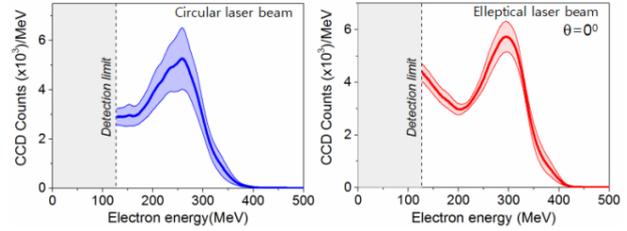

Figure 3. Experimentally measured energy spectra of electron beams produced from LWFA with circular and elliptical focus.

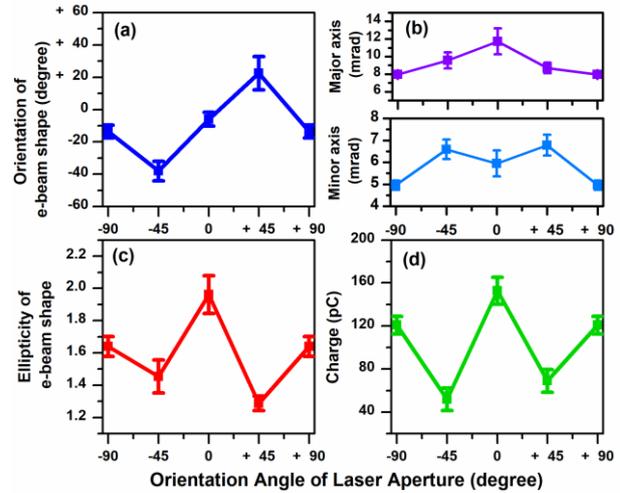

Figure 4. Variation of electron beam parameters with change in orientation of the ellipse shaped laser beam w.r.t. its polarization. For comparison, the electron beam parameters with circular laser beam are (a) orientation: -5.2±3.8, (b) horizontal divergence: -10.2±0.73mrad and vertical divergence: 6.0±0.44mard, (c) ellipticity: 1.7±0.10, and (d) charge: 120±10pC.

The effect of the shaped focal spot on the spatial parameters and the charge of the electron beam is shown in details in Fig. 4. As discussed earlier, the major axis of an electron beam obviously rotated with the rotation of the aperture except for the orientation of 90$^0$. However, the effect of asymmetric wakefield on the transverse profile of the electron beam can be identified from Fig. 4b. The angular divergence along the major axis is highest for the 0$^0$ orientation of the aperture and the lowest for 90$^0$, which highlights the asymmetry (higher strength along the minor axis of the focus) of transverse forces. The variation of different spatial properties of the electron beam shown in Fig. 4 can originate from the combined effect of the drive laser field and the asymmetry in the transverse wakfield caused by the elliptical focal spot. These results clarify the crucial role of transverse forces of the wakefield in controlling LWFA and electron trajectories. More interestingly, the charge of the accelerated electron beam shows a strong dependence on the orientation of the elliptical focal spot w.r.t. the polarization axis. This indicates the delicate influence of the transverse asymmetry of wakefield on electron injection dynamics. It should be noted that the highest charge is injected when the major axis of the elliptical spot is perpendicular to the laser polarization axis and the least charge ($\approx$25% of maximum charge) is injected for $\pm 45^0$. The difference in the charge for $0^0$ and $90^0$ is also noticeable, although the difference of about 25 % is less drastic compared to the cases of $\pm 45^0$.

We performed 3D PIC simulation using the code JoPIC [31] for plasma density of $4.5\times 10^{18}$ cm$^{-3}$ and for the horizontally polarized laser pulses with duration, $\tau$=25fs ($\approx$ half of the plasma wave period) and circular (elliptical) focal spot size 20μm (30μm×20μm) at FWHM. The energy content for both cases is kept constant, i.e. $a_0$=2.7 for the circular focus and 2.2 for the elliptical focus. The plasma density of $4.5\times 10^{18}$ cm$^{-3}$ was chosen to avoid the direct influence of the laser field on accelerating electrons and clearly investigate the effect of shaped cavity alone on the dynamic of LWFA. The plasma density profile is set close to the experimental conditions and the laser focal position is 3 mm where it is behind the plasma target. Simulations clearly show an electron beam with elliptic profile (Fig. 5a) and higher energy (Fig. 5b), even though the initial $a_0$ of the elliptical laser focus is lower than the circular focal spot, as observed in the experiment.

Fig. 5c shows the snap shots at the moment of self-injection in the plasma cavity. The circular focal spot produced transversely circular shaped plasma cavity and the electron injection occured symmetrically, as expected. The formation of transversely elliptic (ellipsoid in 3D) shaped plasma cavity can be clearly seen for the elliptic focal spot. It is interesting to note that the electron injection occurs asymmetrically with the elliptical focal spot wherein more electrons are injected in the vertical plane following the major axis of the laser focus although the transverse wakefield along the major axis is weaker compared to that along the minor axis. The injected electrons acquire larger transverse momentum along the minor axis due to stronger transverse wakefield and produce horizontally ellipse shaped electron beam as shown in Fig. 5(a). The results clearly show that the elliptic-shaped laser can be effectively used to control the trajectories of plasma electrons and confine the self-injection in a desired plane which is otherwise not possible in conventional LWFA.

The simulations also show enhanced electron energy with elliptic laser focus (see Fig. 5b) which can be explained by considering the injection moment. Fig. 5 (d) shows the evolution of the axial wakefield with its propgation distance in the plasma. It shows that sufficient wakefield strength for self-injection develops earlier (dashed line in Fig. 5d) in the case of elleptical focal spot. The earlier injection, therefore, facilitates extended acceleration and higher energy electron beam as shown Fig. 5(b). Both the cases indicate that the injection occurs at about $a_0$ = 3. Even though the elliptical focal spot has lower initial $a_0$, the broader spot size along the major axis produces higher $a_0$ earlier and drives stronger wakefield than that of the circular spot.

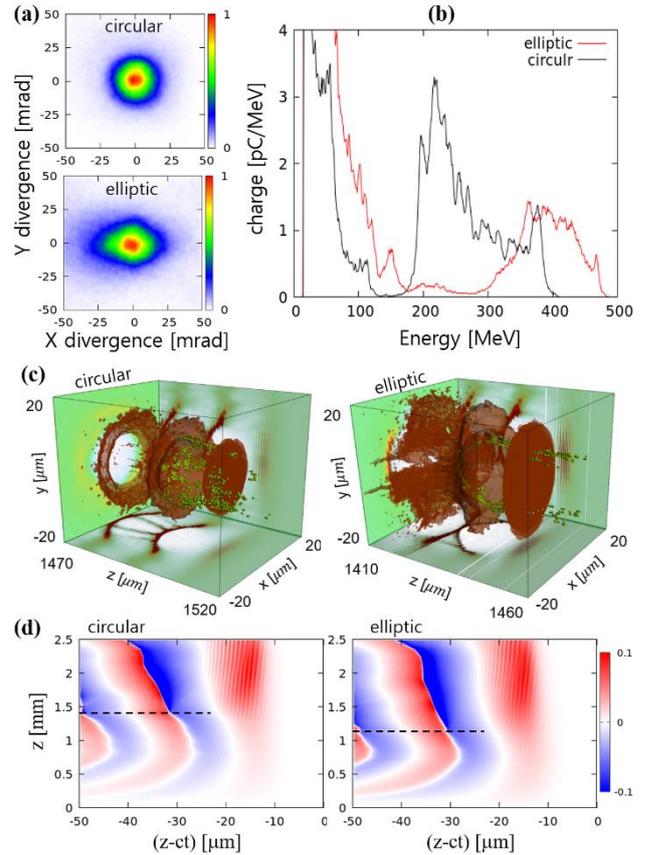

Figure 5. 3D PIC simulation results of LWFA with circular and elliptical focal spots containing the same laser energy: (a) spatial profiles, (b) energy spectra of the electron beams, (c) electron injection in the plasma cavity, and (d) evolution of the wakefield (injection point shown with a dashed horizontal line).

In previous studies, LWFA with asymmetric plasma cavity has been studied with a tilted wavefront [28] or astigmatic aberration [31] of a driving laser pulse to control electron beam direction or to study the angular-momentum evolution of electron beams, respectively. In addition, the generation of an elliptic electron beam along the laser polarization axis, as we observed with circular laser spot, was studied to originate from the direct laser interaction of an accelerated electron beam in the first cavity [29]. In this study, we proved for the first time that the elliptically-shaped laser beams without aberrations or wavefront tilts can be an intrinsic and direct method to control the plasma cavity structure that results in the imprinting of laser profile on the electron beam. The apparent correlation between laser focal spot and electron beam could be verified owing to the aberration-free laser beam obtained with the adaptive control of laser wavefront. We could also decouple the effect of asymmetric plasma cavity and laser polarization by observing the electron beam shape rotating with the elliptic focal shape. Our investigations clearly demonstrated that not only the transverse trajectories of the electrons in the bubble (and consequent modification of the electron beam shape) but also the injection dynamics could be delicately tuned by the choice of the elliptical focal spot and its orientation w.r.t. the laser polarization. Especially, the elliptic plasma cavity controls self-injection and confines betatron oscillations offering a practical knob to tune the electron beam by rotation of the elliptic laser beam. Further, this simple scheme inherently enables emission of polarized betatron x-ray pulses from LWFA without the need for ionization induced injection [32].

In conclusion, we demonstrated that the transverse shaping of a laser focal spot improved the stability and energy of electron beams produced by effectively controlling the transverse shape of plasma cavities in LWFA. The cavity shaping and its rotation provided control on electron injection dynamics and the consequent variation of accelerated electron charge, energy, and the plane of betatron oscillations. With adaptive optics available at high power laser facilities, in situ focal spot shaping can be routinely done in a feedback loop to accomplish shaped electron and polarized x-ray beams at target location for future research and applications of LWFA.

The authors would like to acknowledge the support by the Institute for Basic Science (IBS-R012-D1) and by the Research on Advanced Optical Science and Technology grant funded by GIST.

[1] T. Tajima and J. M. Dawson, Phys. Rev. Lett. **43**, 267 (1979).
[2] E. Esarey, C. B. Schroeder, and W. P. Leemans, Rev. Mod. Phys. **81**, 1229 (2009).
[3] C. Joshi, IEEE Trans. Plasma Sci. **49**, 3134 (2017).
[4] C. B. Schroeder, E. Esarey, C. G. R. Geddes, C. Benedetti and W. P. Leemans, Phys. Rev. Special Topics **13**, 101301 (2010).
[5] M. E. Couprie, A. Loulergue, M. Labat, R. Lehe and V. Malka, J. Phys. B: At. Mol. Opt. Phys. **47**, 234001 (2014).
[6] H. T. Kim, J. H. Shin, C. Aniculaesei, B. S. Rao, V. B. Pathak, M. H. Cho, C. Hojbota, S. K. Lee, J. H. Sung, H. W. Lee, J. W. Yoon, K. Nakajima and C. H. Nam, The 46th European Physical Society Conference on Plasma Physics, p. **11.**204 (2019).
[7] A. J. Gonsalve et al., Phys. Rev. Lett. **122**, 084801 (2019).
[8] M. H. Cho, V. B. Pathak, H. T. Kim, and C. H. Nam, Scientific reports **8**, 16924 (2018).
[9] J. Faure, C. Rechatin, A. Norlin, A. Lifschitz, Y. Glinec, and V. Malka, Nature **444**, 737 (2006).
[10] A. Pak, K. A. Marsh, S. F. Martins, W. Lu, W. B. Mori, and C. Joshi, Phys. Rev. Lett. **104**, 025003 (2010).
[11] B. B. Pollock et al., Phys. Rev. Lett. **107**, 045001 (2011).
[12] A. J. Gonsalves et al., Nat. Phys. **7**, 862 (2011).
[13] B. S. Rao et al., Appl. Phys. Lett. **102**, 231108 (2013).
[14] A. Buck et al., Phys. Rev. Lett. **110**, 185006 (2013).
[15] R.Lehe et al., Phys. Rev. Lett. **111,** 085005 (2013).
[16] B. S. Rao et al., Phys. Rev. ST Accel. Beams **17**, 011301 (2014).
[17] J. Wenz et al., Nat. Photonics **13**, 263 (2019).
[18] J. Luo, M. Chen, M. Zeng, J. Vieira, L. L. Yu, S. M. Weng, L. O. Silva, D. A. Jaroszynski, Z. M. Sheng and J. Zhang, Sci. Rep. **6**, 29101 (2016).
[19] S. Corde, K. Ta Phuoc, R. Fitour, J. Faure, A. Tafzi, J. P. Goddet, V. Malka, and A. Rousse, Phys. Rev. Lett. **107**, 255003 (2011)
[20] S. P. D. Mangles et al., Appl. Phys. Lett. **95**, 181106 (2009).
[21] S. Corde, K. Ta Phuoc, G. Lambert, R. Fitour, V. Malka, A. Rousse A. Beck and E. Lefebvre, Rev. Mod. Phys. **85**, 1 (2013).
[22] A. Pukhov and J. Meyer-ter-Vehn, Appl. Phys. B Lasers Opt. **74**, 355–361 (2002).
[23] W. Lu, C. Huang, M. Zhou, W. B. Mori, and T. Katsouleas, Phys. Rev. Lett. **96**, 165002 (2006).
[24] W. Lu, M. Tzoufras, C. Joshi, F. S. Tsung, W. B. Mori, J. Vieira, R. A. Fonseca, and L. O. Silva, Phys. Rev. ST Accel. Beams **10**, 061301 (2007).
[25] I. Kostyukov, E. Nerush, A. Pukhov, and V. Seredov, Phys. Rev. Lett. **103**, 175003 (2009)
[26] S. Kalmykov, S. A. Yi, V. Khudik, and G. Shvets, Phys. Rev. Lett. **103**, 135004 (2009).
[27] Y. Glinec, J. Faure, A. Lifschitz, J. M. Vieira, R. A. Fonseca, L. O. Silva, and V. Malka, Europhys. Lett. **81**, 64001 (2008).
[28] A. Popp et al., Phys. Rev. Lett. **105**, 215001 (2010).
[29] C. Thaury, E. Guillaume, S. Corde, R. Lehe, M. Le Bouteiller, K. Ta Phuoc, X. Davoine, J. M. Rax, A. Rousse, and V. Malka, Phys. Rev. Lett. **111**, 135002 (2013).
[30] C. A. Peters, Statistics for Analysis of Experimental Data, Environmental Engineering Processes Laboratory Manual, AEESP, Champaign, IL (2001).
[31] S. P. D. Mangles et al., Phys. Rev. Lett. **96**, 215001 (2006).
[32] Andreas Döpp et al., Light: Science & Applications **6**, e17086 (2017).